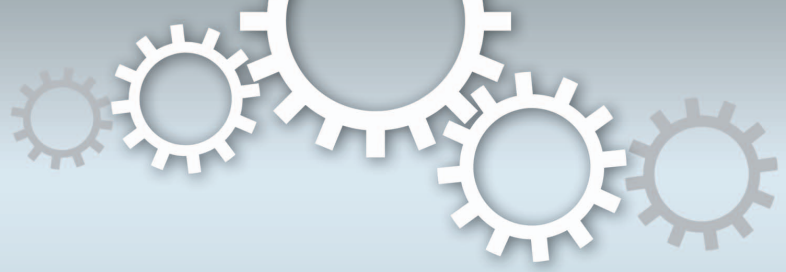

OPEN

# Bidirectional selection between two classes in complex social networks

Bin Zhou[1,2], Zhe He[1], Luo-Luo Jiang[3], Nian-Xin Wang[2] & Bing-Hong Wang[1,3]

[1]Department of Modern Physics, University of Science and Technology of China, Hefei, 230026, China, [2]School of Economics and Management, Jiangsu University of Science and Technology, Zhenjiang, 212003, China, [3]College of Physics and Electronic Information Engineering, Wenzhou University, Wenzhou, 325035, China.

The bidirectional selection between two classes widely emerges in various social lives, such as commercial trading and mate choosing. Until now, the discussions on bidirectional selection in structured human society are quite limited. We demonstrated theoretically that the rate of successfully matching is affected greatly by individuals' neighborhoods in social networks, regardless of the type of networks. Furthermore, it is found that the high average degree of networks contributes to increasing rates of successful matches. The matching performance in different types of networks has been quantitatively investigated, revealing that the small-world networks reinforces the matching rate more than scale-free networks at given average degree. In addition, our analysis is consistent with the modeling result, which provides the theoretical understanding of underlying mechanisms of matching in complex networks.

Understanding the pattern of human activities has been received growing attention due to it's important practical applications from traffic management to epidemic control[1–4]. Several mechanisms with individual activities have been discovered based on statistics of huge amounts of data on human behaviors, such as queueing theory and adaptive interest[5–7]. However, mechanisms behind human activities with interacting individuals are far from well understood because of complex population structures which can be described by complex networks[8–10]. Apart from the statistic characteristics of human dynamics in space and time interval, abundant researches have focused on comprehending human activities in social networks such as making friends where people in the same class with similar feature are more likely to be friends. There are also common phenomena of seeking social partners belonging two classes in bipartite populations[11,12], such as mate choosing between men and women, commercial trading between buyers and sellers.

The seeking processes which may be the base of building many social relationships can be described by matching model[13]. Individuals are generally divided into two classes according to their natural status. Then they observe features of others belonging to the other class, and finally decide whether select the individual as a social partner. Although characters of individuals are too complex to be quantitatively described in bidirectional selection systems, personal quality and economic status can be viewed as the main characters of individuals[14,15]. Zhang and his collaborators solve the bipartite matching problem in the framework of economic markets, finding that partial information and bounded ratio-nality contribute to satisfied and stable matches[16,17]. Besides characters of individuals, the matching processes is also affected by structure of social networks[18–20]. Since social networks have emerged some common characteristics, such as small-world phenomena, scale-free properties with power-law degree distributions[21,22]. Questions naturally arise how properties of social networks affect matching processes, and what kind of the property improves the matching performance of networks. To answer these questions, a bipartite network is reconstructed from the original networks[23–25], where only connected nodes satisfying successfully matching conditions and their links are reserved. This allows us to investigate the bidirectional matching processes with mathematical analysis and computer simulation.

In this paper, we researched the matching problem of two classes in the framework of complex networks. The analytical solution for the rate of successfully matching rate is presented, which is consistent with our simulation results of matching processes on social networks. It is observed that properties of networks greatly impact matching performance of networks, and the small-world effect improve rate of successfully matching more than scale-free properties. In addition, the small-world effect on matching performance of networks was quantitatively investigated with different rewiring rate in the small world network.

## Results

For the given network, M nodes belong to the class A and N nodes belong to the class B (see Methods). After the characteristic state of each node is determined, for a node in the given network, only the neighbor nodes which can successfully match with the node are valuable to the node. Thus the original network is reconstructed as a bipartite network where only connected nodes satisfying successfully matching conditions and their links are reserved, as shown in top panels of Fig. 1 (a). In this way, we get a new bipartite network with $m$ ($m \leq M$) nodes belonging to the class A and $n$ ($n \leq N$) nodes belonging to class B, where any two connected nodes satisfy conditions of successfully matching. In the new bipartite network, $k_i$ is the degree of the $i$th node in class A, and $k_h$ denotes the degree of the $h$th node in class B connected to the $i$th node of class A. In order to get the probability on successfully matching of the $i$th node, we can firstly calculate the probability on unsuccessfully matching of the $i$th node. Because the degree of the $h$th node in class B is $k_h$, the probability that the $h$th node can successfully match with the $i$th node is $1/k_h$. Therefore, the probability that the $h$th node can unsuccessfully match with the $i$th node is $1 - 1/k_h$. So the probability that the $i$th node in class A can not successfully match with all neighbor nodes is

$$\prod_{h=1}^{k_i}\left(1-\frac{1}{k_h}\right). \quad (1)$$

The probability that the $i$th node in class A can be successfully matched is

$$1-\prod_{h=1}^{k_i}\left(1-\frac{1}{k_h}\right). \quad (2)$$

If $k_i$ denotes the degree of node $i$ in class B, $k_h$ presents the degree of $h$th node in class A connected to the $i$ node of class B, and above equations also describe nodes in class B. As shown in Fig. 1 (b), the the probability that node $i$ can not match with its neighbors is 0.0 for node $A_1$ of class A, 0.5 for node $B_1$ of class B, and 0.5 for node $B_2$ of class B (The three nodes are shown in the bipartite network of top panels in Fig. 1). If there are $m$ nodes belonging to the class A, the expectation $E$ of the total number of nodes in domain A matched successfully is

$$E(m,n,\mu)_A = m - \sum_{i=1}^{m}\prod_{h=1}^{k_i}\left(1-\frac{1}{k_h}\right), \quad (3)$$

where the $\mu$ denotes the number of types for nodes' characters. Similarly, the expectation $E$ of the total number of nodes in class B matched successfully is

$$E(m,n,\mu)_B = n - \sum_{j=1}^{n}\prod_{h=1}^{k_j}\left(1-\frac{1}{k_h}\right). \quad (4)$$

Because the matching between the two classes is one-to-one, $E(m, n, \mu)_A = E(m, n, \mu)_B$. Therefore, for a given network with M, N and $\mu$, the expectation $E$ about the total number of successfully matching pairs in the model is

$$E(M,N,\mu) = m - \sum_{i=1}^{m}\prod_{h=1}^{k_i}\left(1-\frac{1}{k_h}\right) = n - \sum_{j=1}^{n}\prod_{h=1}^{k_j}\left(1-\frac{1}{k_h}\right). \quad (5)$$

According to (5), we can get

$$2E(M,N,\mu) = m - \sum_{i=1}^{m}\prod_{h=1}^{k_i}\left(1-\frac{1}{k_h}\right) + n - \sum_{j=1}^{n}\prod_{h=1}^{k_j}\left(1-\frac{1}{k_h}\right). \quad (6)$$

Further, we define the average successful matching rate of networks from the equation (6):

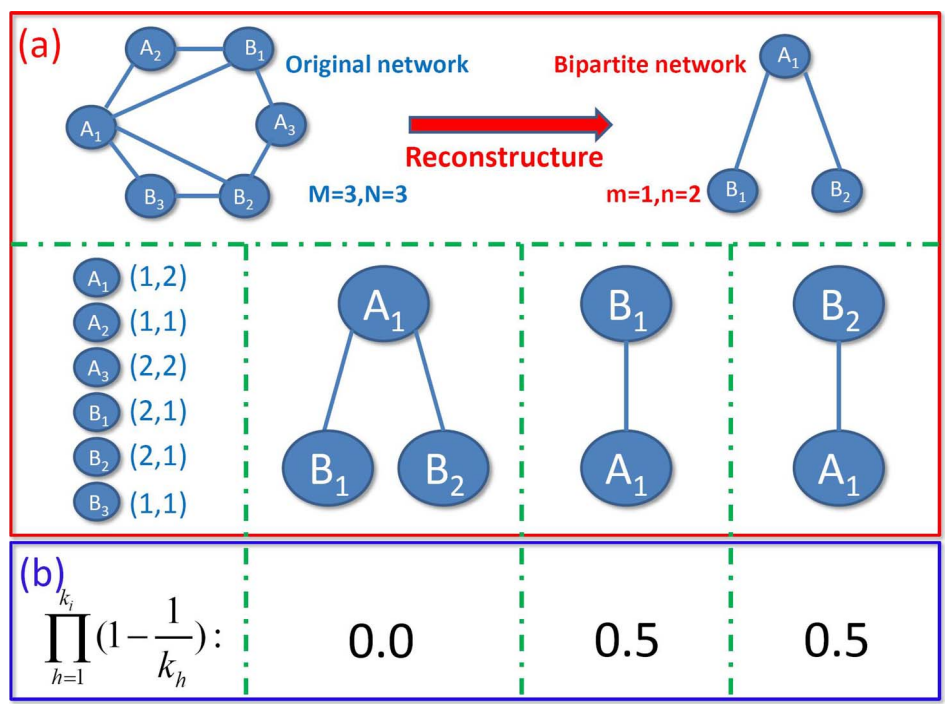

**Figure 1 | Illustration of the matching process.** Upper panel (a) gives a simple example how the bipartite network is reconstructed, where node $A_1$ in set A has two neighbors belonging to set B where $B_1$ and $B_2$ are matched with $A_1$, $m = 1$ and $n = 2$. Compared to the original networks, only links of matched nodes are reserved in the reconstructed network. Here, $A_1$ with character of (1,2), $B_1$ with character of (2,1), and $B_2$ with character of (2,1) are reserved. $\prod_{h=1}^{k_i}\left(1-\frac{1}{k_h}\right)$ denotes the probability that node $i$ can not match with its neighbors, and the value is 0.0 for $A_1$, 0.5 for $B_1$, and 0.5 for $B_2$, as shown in panel (b). Thus the expectation to match successfully for the set A is $E_A = m - \sum_{i=1}^{m}\prod_{h=1}^{k_i}\left(1-\frac{1}{k_h}\right) = 1.0$, and the successfully matching expectation of the set B is calculated as $E_B = n - \sum_{i=1}^{n}\prod_{h=1}^{k_i}\left(1-\frac{1}{k_h}\right) = 1.0$.

$$\Lambda = \frac{2E(M,N,\mu)}{M+N}$$

$$= \frac{m - \sum_{i=1}^{m} \prod_{h=1}^{k_i}\left(1-\frac{1}{k_h}\right) + n - \sum_{j=1}^{n} \prod_{h=1}^{k_i}\left(1-\frac{1}{k_h}\right)}{M+N}, \quad (7)$$

where $M + N$ represents the total population of two classes in the network, and the range of $\Lambda$ is from 0 to 1. Therefore, $\Lambda$ can quantify the matching performance of a network, and the large value of $\Lambda$ reflects high matching performance of the network.

To investigate the effects of population structures on the matching rates of networks, analytical results are performed from above equations, as shown in top panels of Fig. 2. Without loss of generality, $\mu$ is fixed as 2 in the analysis, and four types of networks are applied to model different structured population (see Methods). One can find that there exists an optimal value of $\alpha$ ($\alpha_o \approx 0.5$) to enhance matching performance of networks, revealing that balanced population between class $A$ and $B$ plays an important role in matching performance for the four networks. In addition, the average degree of networks $K$ affects $\Lambda$ greatly nearby the $\alpha_o$, and a larger average degree of networks induces better matching performance of networks.

To confirm analytical results, we performed simulations of matching process on regular networks, small-world networks, random networks and scale-free networks respectively, as shown in the bottom panels of Fig. 2. In simulation, firstly of all, nodes are divided to the class $A$ with probability $\alpha$ ($0 \leq \alpha \leq 1$), and with probability $1 - \alpha$ belong to class $B$. Then, the state of each node is randomly assigned a kind of character from $\Omega$ following the uniform distribution and $\mu$ is also fixed as 2 in the simulation. For example, the characters of node $i$ are labeled as $(c_{A_i}, s_{A_i})$ where $c_{A_i}$ is node $i$'s own character and $s_{A_i}$ represents the character the node $i$ attempts to select. It is found that our simulation results are consistent with analysis, and both results show peaks of the $\Lambda$ all appearing at around $\alpha = 0.5$ where the peak of $\Lambda$ on the scale-free network is the lowest. We therefor focus on the matching performance of different average degree $K$ at $\alpha = 0.5$. Fig. 3 indicates the rate of successfully matching increases with enhancing of average degree of networks, and the $\Lambda$ value of scale-free network is minimum at the same value of the average degree $K$ compared to other networks. On the other hand, the successfully matching rate $\Lambda$ of small-world networks is the maximum, reflecting that structures of small-world are more conducive to matching process than structures of scale-free networks.

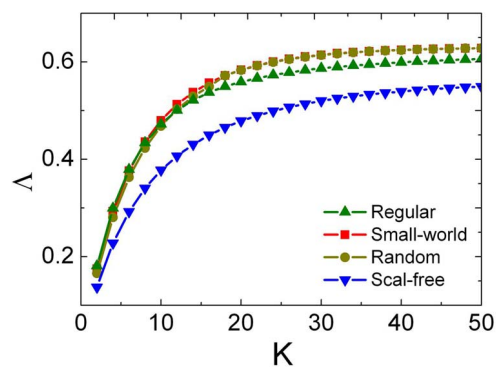

**Figure 3** | **Matching performance of different networks as a function of average degree $K$.** The total population $M + N = 1000$, and $\mu = 2$. For WS small-world networks, the rewiring rate $\beta$ is fixed as 0.5.

Since structures of small-world networks enhance the matching performance of networks greatly, we focused on the matching process on small-world networks with different rewiring probability $\beta$, as shown in Fig. 4. It is found that $\Lambda$ monotonously decreases with increasing of $\beta$ in the condition of $\alpha = 0.1$ where individuals in class $B$ are much more than that of class $A$. When the balanced population between class $A$ and class $B$ is achieved, i.e. the value of $\alpha$ is nearby 0.5, there exists an optimal value of rewiring rate $\beta$ to induce the highest rate of successfully matching for small-world networks, such as Fig. 4(b) and Fig. 4(c). It indicates that appropriate rewiring links contribute to improve matching performance of small-world networks with balanced population between class $A$ and class $B$. It is worthy mentioning that intrinsic results do not change for different values of $\mu$, such as $\mu = 2$ and $\mu = 5$.

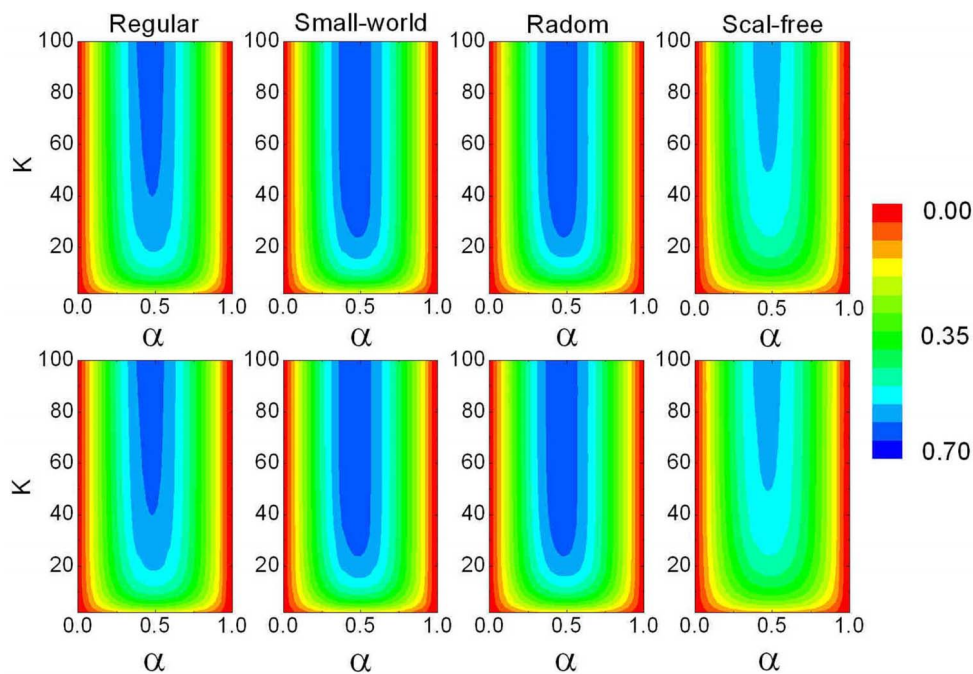

**Figure 2** | **Matching performance of different networks in $K - \alpha$ parameter phase space.** Top and bottom panels show analytical and simulation results of matching performance for different networks respectively. The total population $M + N = 1000$, and $\mu = 2$. For WS small-world networks, the rewiring rate $\beta$ is fixed as 0.5.

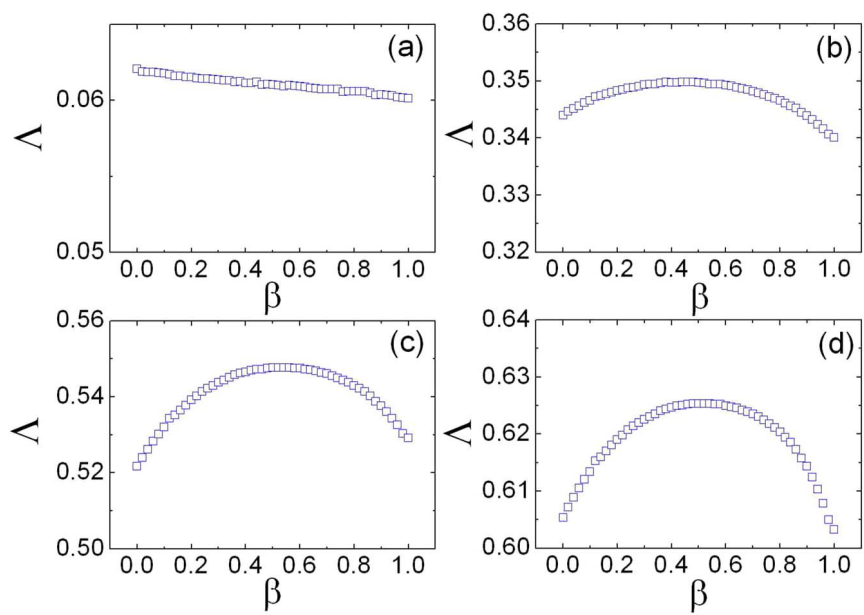

**Figure 4 | Matching performance as a function of the rewiring rate $\beta$ for WS small-world networks.** (a) for $K = 4$ and $\alpha = 0.1$, (b) for $K = 20$ and $\alpha = 0.25$, (c) for $K = 50$ and $\alpha = 0.35$. (d) for $K = 100$ and $\alpha = 0.5$. The total population $M + N = 1000$, and $\mu = 3$.

In order to study effects of population size on the rate of successfully matching, we conducted simulation of matching processes on random networks with the average degree of $K = 1$, $K = 2$, $K = 3$, and $K = 4$ respectively, as shown in Fig. 5. In the case of $K = 1$ with $m = n = 1$, the probability of successfully matching between the two connected nodes belonging two classes is about 0.125, which is consistent with the mathematical analysis result. If the two connected nodes match with each other successfully, they must simultaneously satisfy the two conditions. Firstly, the two connected nodes belong to different classes $A$ and $B$. The second, $c_{A_i} = s_{B_j}$ and $s_{A_i} = c_{B_j}$, where $i$ and $j$ represent the two connected nodes. The probability of satisfying the first condition is 1/2 and the probability of satisfying the second condition is 1/4, so the probability of successfully matching between the two connected nodes is 1/8. Limited by average degree of networks, the number of total populations in the four case of $K = 1$, $K = 2$, $K = 3$, and $K = 4$, starts from 2, 3, 4, and 5 respectively. One can find that successfully matching rate $\Lambda$ decrease with increasing of $M + N$ when the number of total populations is lower than 10, while the value of $\Lambda$ tends to a stable value for $M + N > 10$.

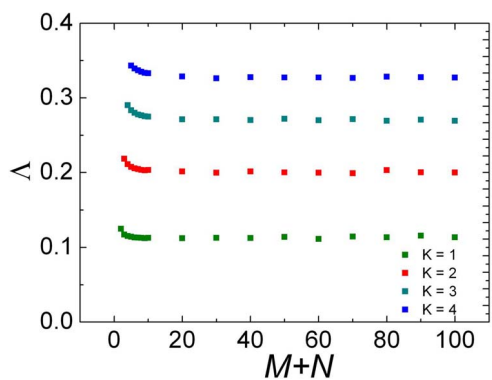

**Figure 5 | Matching performance as a function of total population for ER random networks with different average degree.** To avoid repeating links, the total population $M + N$ for average degree $K = 1$, $K = 2$, $K = 3$, and $K = 4$, starts from 2, 3, 4, and 5 respectively. $\mu = 2$ and $\alpha = 0.5$.

## Discussion

Although our results are obtained from mathematical analysis and computer simulation, there are some human subject experiments which support our conclusions. For example, in experiments on matching behavior[20], human subjects are connected on virtual complex networks with the interface of computers, including preferential attachment and small-world networks. Different from our model where all individuals are divided into two classes, participants in their experiments belong to one class to march as a single pair. In particularly, subjects in the human experiments are able to propose to match with a neighbor and accept a proposal from a neighbor, which is similar to matching process of our model. The experimental results show that the matching performance of small-world networks is better than that of preferential attachment networks, which are consistent with our conclusions. In addition, the similar observation is also obtained from the experimental data of the coloring games performed by Kearns et al.[18], where preferential attachment networks lead to worse performance than small-worlds networks.

It is worth mentioning that our approach is also suitable for the condition of fully connected networks where the average degree of networks depends on the size of networks. In this case, the matching performance is determined by the size of networks, and the larger networks lead to the higher successfully matching rate, which is consistent with the result of real data[13]. Compared to the previous work[13], the current model and analytical solutions can be used to solve the matching problem in complex networks, thus the illustrations of matching processes has been extended in more general situations. In particularly, the matching process on small-world networks with different rewiring probability was studied in details, because the structures of small-world networks obviously enhance the matching performance of networks.

Summarizing, we have studied the bidirectional selection system on complex networks where nodes are occupied by individuals in two classes. The average matching rate is proposed to evaluate the successfully matching performance of networks. It found that high average degree of nodes and balanced population between the two classes contributes to enhance the matching performance of networks, and our analysis is consistent with the simulation results. We also observed that the small-world networks perform better than scale-free networks at a given average degree. Our approach to restructure the bipartite network may be also applied to spreading dynamics of information and diseases in bipartite populations[26–28], where some social partners would be successful in matching but others not. There are also future application of our research in co-volution of matching dynamics and social network structures[29,30].

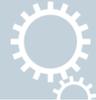

## Methods

**The matching model in structured population.** Our model of the bidirectional selection on complex networks is stated as follows:

i) For a given network, nodes are occupied by individuals who are divided into class $A$ with $M$ individuals and class $B$ with $N$ individuals. Therefore, the total population in the networks is $M + N$.

ii) The $i$th node in class $A$ (or $j$th node in class $B$) has its own character denoted by $c_{A_i}$ (or $c_{B_j}$). Correspondingly, the character the $i$th node attempts to select is denoted by $s_{A_i}$ (or the $j$th node attempts to select is denoted by $s_{B_j}$). Therefore, the characteristic state of the $i$th node (or the $j$th node) is denoted by $(c_{A_i}, s_{A_i})$ (or $(c_{B_j}, s_{B_j})$) in the bidirectional selection.

iii) The nodes' characters are labeled as integers. Assume the characters have $\mu$ types, i.e. $c_{A_i}, c_{B_j}, s_{A_i}, s_{B_j} \in \Omega$, $\Omega = \{1, 2, \ldots, \mu\}$, $i \in \{1, 2, \ldots, M\}$, $j \in \{1, 2, \ldots, N\}$. Therefore, the characteristic state of a node can be denoted by $(\mu, 1)$, $(2, 1)$, $(\mu, 2)$, etc.

iv) In the network, the conditions of successfully matching for two connected nodes $A_i$ and $B_j$ are $c_{A_i} = s_{B_j}$ and $s_{A_i} = c_{B_j}$. That is, when node $A_i$'s character meets node $B_j$'s requirement and *vice versa*, two connected nodes $A_i$ and $B_j$ have a successful matching.

To model the structure of population, regular networks, small-world networks, random network and scale-free networks are conducted as follows:

1) Regular networks: Starting from a regular ring lattice with $M + N$ vertices with $K$ edges per vertex, each vertex connected to its $K$ nearest neighbors by undirected links[21].

2) Small world network: Starting from regular network with degree of $K$, we randomly choose a vertex and its edge, then rewiring the link to an randomly selected node with probability $\beta$, until each edge in the original regular networks has been considered once[21].

3) Random network: Starting from a regular network with $M + N$ nodes, we connect any two nodes with the probability $K/(M + N - 1)$, where the $K$ is the average degree of network[31].

4) Scale-free network: First of all, a globally coupled network with $K + 1$ nodes is built, where $K$ is the average degree of a network[22]. Then, the network grows with preferential attachment process with the probability that a new node will be connected to the node $i$ is proportional to the degree of the node. The network keep growing until the size of the network is up to $M + N$.

**Simulations.** For a given network, in the simulation trial of the model, nodes belong to the class $A$ with probability $\alpha$, and belong to class $B$ with probability $1 - \alpha$. The number of nodes in class $A$ and class $B$ are $M$ and $N$ respectively. Therefore, there are $M + N$ nodes in the network. For the characters of each node, the $c_{A_i}, s_{A_i}, c_{B_j}, s_{B_j}$ are randomly assigned a kind of character from set $\Omega$ with $\mu$ types of characters. A bipartite network is reconstructed from the given network, where only the matched nodes and links of matched nodes are reserved. Then a new bipartite network is generated with $m$ ($m \leq M$) nodes belonging to class $A$ and $n$ ($n \leq N$) nodes belonging to class $B$. In the new bipartite network, a node and one of its neighboring nodes are randomly chosen. The two nodes are determined as a pair, meaning the two nodes are matched successfully. Every node can be only chosen one time at most. The matching process is repeated until no pair can be determined any more. In this way, we can calculate how many nodes are matched successfully, and get the rate of successfully matching for the whole population.

### Acknowledgments
We thank Ming Li for helpful comments and suggestions. This work is supported by: The National Natural Science Foundation of China (Grant Nos. 61203145, 11275186, 91024026, 71331003, 71271104 and 61403421) and The Open Funding Programme of Joint Laboratory of Flight Vehicle Ocean-based Measurement and Control(Grant Nos. FOM2014OF001).

### Author contributions
B.Z. designed and performed the research. B.Z., Z.H. and L.-L.J. analyzed the result. B.Z., L.-L.J., Z.H., N.-X.W. and B.-H.W. wrote the paper.

### Additional information
**Competing financial interests:** The authors declare no competing financial interests.

**How to cite this article:** Zhou, B., He, Z., Jiang, L.-L., Wang, N.-X. & Wang, B.-H. Bidirectional selection between two classes in complex social networks. *Sci. Rep.* **4**, 7577; DOI:10.1038/srep07577 (2014).